\documentclass{article}

\usepackage{PRIMEarxiv}
\usepackage[utf8]{inputenc} 
\usepackage[T1]{fontenc}    
\usepackage{hyperref}       
\usepackage{url}            
\usepackage{booktabs}       
\usepackage{amsfonts}       
\usepackage{nicefrac}       
\usepackage{microtype}      
\usepackage{lipsum}         
\usepackage{fancyhdr}       
\usepackage{graphicx}       
\graphicspath{{media/}}     
\usepackage{amsmath}

\title{Improving Pedestrian Safety at Intersections Using Probabilistic Models and Monte Carlo Simulations
\thanks{\textit{\underline{Citation}}: 
\textbf{Alben Rome Bagabaldo and Jürgen Hackl. Improving Pedestrian Safety at Intersections Using Probabilistic Models and Monte Carlo Simulations. Pages..., DOI:...}} 
}

\author{
  Alben Rome Bagabaldo, Jürgen Hackl \\
  The Department of Civil and Environmental Engineering \\
  Princeton University \\
  Princeton, New Jersey\\
  \texttt{\{alben, hackl\}@princeton.edu} \\
}

\begin{document}
\maketitle

\begin{abstract}
National Highway Traffic Safety Administration reported 7,345 pedestrian fatalities in the United States in 2022, making pedestrian safety a pressing issue in urban mobility. This study presents a novel probabilistic simulation framework integrating dynamic pedestrian crossing models and Monte Carlo simulations to evaluate safety under varying traffic conditions. The framework captures key influences on pedestrian decisions, such as traffic light states, vehicle proximity, and waiting times, while employing the Intelligent Driver Model (IDM) to simulate realistic vehicle dynamics. Results from 500 trials show that pedestrians avoid crossing during green lights, reducing collision risks, while shorter waiting times during red lights encourage safer crossings. The risk is heightened during yellow lights, especially with nearby vehicles. This research emphasizes the importance of adaptive traffic control measures, such as pedestrian-triggered signals and enhanced traffic light timing, to mitigate risks and prioritize pedestrian safety. By modeling realistic interactions between pedestrians and vehicles, the study offers insights for designing safer and more sustainable urban intersections.
\end{abstract}


\section{Introduction}

Road infrastructures are fundamental to urban transportation systems, facilitating the interaction of diverse road users, including vehicles, cyclists, pedestrians, and other vulnerable users. However, intersections are particularly complex environments to manage, often serving as bottlenecks and hotspots for traffic accidents. Pedestrian injuries comprise a substantial portion of road traffic fatalities, underscoring the disproportionate risks faced by vulnerable road users \cite{who2023}. National Highway Traffic Safety Administration reported 7,345 pedestrian fatalities in the United States in 2022 \cite{nhtsa2022}. These fatalities frequently occur at intersections, where conflicting movements and insufficient safety measures increase accident risks \cite{gitelmanEtAl2020, mahmoudiEtAl2022}.

The challenges of intersection management are multifaceted, requiring solutions that account for dynamic interactions between road users and their environment. Key factors such as traffic light states, pedestrian waiting times, vehicle proximity, and environmental conditions significantly influence pedestrian safety outcomes. Traditional approaches, such as fixed signal timings, are often inadequate in addressing the complexities of modern urban traffic systems. Advanced computational models, such as stochastic hybrid systems, offer promising avenues for improving intersection safety by dynamically adapting traffic light policies based on real-time traffic conditions, optimizing waiting times for both vehicles and pedestrians \cite{chenCassandras2024}.

\textit{Contribution: } This study presents a novel pedestrian crossing simulation framework that integrates probabilistic reasoning and statistical distributions to model pedestrian and vehicle speeds. The framework evaluates pedestrian decisions and collision risks under various traffic conditions. The Intelligent Driver Model (IDM) is incorporated to capture realistic car-following behavior. Monte Carlo simulations strengthen the model's robustness by replicating real-world scenarios to assess its applicability in enhancing pedestrian safety. The insights gained aim to inform urban planners and traffic engineers, providing actionable recommendations for safer and more sustainable intersection designs.

The organization of this paper is as follows: The Methodology in section \ref{sec:methodology} details the design of the Simulation Framework, Vehicle Dynamics and Movement Logic, Dynamic Modeling of Pedestrian Crossing Behavior, and Monte Carlo Simulations. Section \ref{sec:results} - Results and Discussion highlights the key insights derived from the simulation experiments, emphasizing their practical implications for intersection management. Finally, Section \ref{sec:conclusion} 
 - Conclusion summarizes the study's contributions, discusses its practical implications, and suggests potential directions for future research in pedestrian safety and urban mobility.

\section{Methodology}
\label{sec:methodology}

\begin{figure}[ht]
  \centering
  \includegraphics[scale=.4]{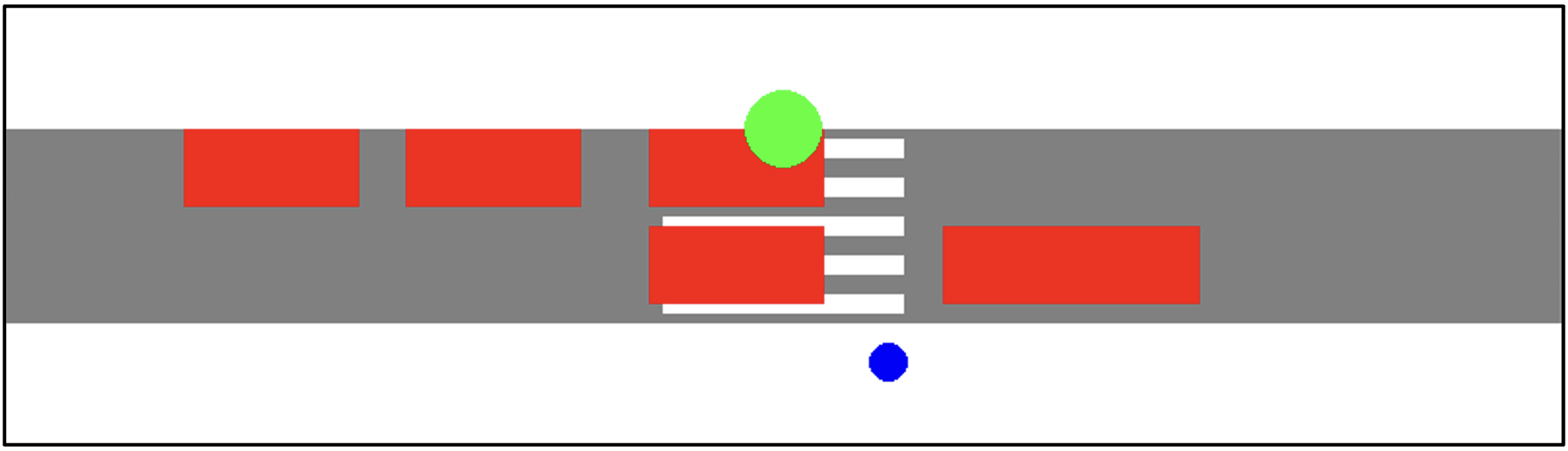}
  \caption{\textbf{Screenshot of the simulation setup created using PyGame.} The blue circle represents the pedestrian, the red rectangles represent vehicles, and the green circle represents the traffic light, which transitions between green, yellow, and red states.}
  \label{fig:fig1}
\end{figure}

The simulation was developed using PyGame \cite{pygame2024}, a free and open-source Python library for creating multimedia applications. It was selected for its simplicity, accessibility, and flexibility, which facilitate rapid prototyping of simulation environments. Leveraging PyGame’s event handling, rendering, and animation features, the pedestrian crossing simulation was constructed as a visually intuitive representation. Figure \ref{fig:fig1} illustrates a screenshot of the simulation setup, which includes dynamic objects and visual cues designed to mimic real-world intersection conditions.

\subsection{Simulation Framework}

The simulation framework comprises three core components: \textit{Pedestrians}, \textit{Vehicles}, and \textit{Traffic Lights}. These objects interact dynamically to replicate traffic scenarios and evaluate pedestrian crossing safety. Each component is designed to model realistic behaviors and integrate seamlessly into the simulation environment.

Pedestrian behavior is modeled using a probabilistic model (see Section \ref{sec:pedestrian-behavior} - Dynamic Modeling of Pedestrian Crossing Behavior), which determines crossing decisions based on three key factors: waiting time, vehicle distance, and traffic light state. Pedestrians begin at random positions on either side of the crosswalk and move vertically, up or down, to cross. Walking speeds in the simulation are sampled from a normal distribution, in line with findings by Chandra and Bharti \cite{chandraBharti2013}, who demonstrated that pedestrian walking and crossing speeds generally follow a normal distribution. At each trial, the dynamic model evaluates whether the pedestrian will cross or wait, balancing probabilistic reasoning with real-time interactions to ensure realistic pedestrian behavior in the simulation.

The vehicles in the simulation are programmed to follow the Intelligent Driver Model (IDM) \cite{treiberKesting2013}, ensuring realistic acceleration, deceleration, and interaction with other vehicles. They move within designated lanes, traveling either left-to-right or right-to-left. Vehicle speed is determined based on the distance to the vehicle ahead, as calculated using IDM, and the state of the traffic light. To prevent collisions, vehicles maintain safe stopping distances and respond appropriately to pedestrians at crosswalks. Additionally, vehicles are spawned with varying speeds, sampled from a uniform distribution.

The traffic light governs the flow of vehicles and influences pedestrian decisions. The light cycles through green, yellow, and red states based on predefined timing intervals. The light's state is displayed using color-coded indicators above the crosswalk. The traffic light states are also expected to directly affect vehicle speeds and pedestrian crossing probabilities, providing a central mechanism for coordinating interactions.

Meanwhile, the simulation environment represents a two-lane road intersecting with a crosswalk at its center as shown in Figure \ref{fig:fig1}. The crosswalk spans the entire width of the intersection and features visual stripes to guide pedestrians. Two opposing lanes allow vehicles to traverse the intersection and dynamically interact with the crosswalk. Distances between objects, such as vehicle-to-pedestrian proximity and crosswalk width, are measured in pixels and converted to real-world scales using a pixels-per-meter (PPM) conversion. Variables like speed and acceleration are adjusted using a combination of frames per second (FPS) and PPM for accurate scaling.

This framework forms the foundation for the Monte Carlo simulations, discussed in Section \ref{sec:monte-carlo}, which evaluates pedestrian safety and collision probabilities across diverse scenarios.

\subsection{Vehicle Dynamics and Movement Logic}
\label{sec:vehicle-dynamics}

Vehicle dynamics are modeled using realistic traffic behavior principles to ensure accurate interactions with other vehicles, pedestrians, and traffic lights. Each vehicle is assigned a lane corresponding to its direction of travel, spawned at random positions off-screen to ensure a steady flow. Initial speeds are sampled from a uniform distribution (\textit{e.g.,} 10-15 $m/s$). Vehicles are parameterized with safe distance, maximum acceleration, and comfortable braking deceleration values as defined by the IDM. Vehicle acceleration is computed using the IDM formula in equation \ref{eqn:acceleration}.

\begin{equation}
\label{eqn:acceleration}
a = a_{\text{max}} \left(1 - \left(\frac{v}{v_0}\right)^\delta - \left(\frac{s^*}{s}\right)^2 \right)
\end{equation}

where \(v\) is the current speed, \(v_0\) is the desired speed, \(a_{\text{max}}\) is the maximum acceleration, \(s\) is the gap to the vehicle in front, and \(s^*\) is the desired dynamic gap as described in equation \ref{eqn:gap}.

\begin{equation}
\label{eqn:gap}
s^* = s_0 + vT + \frac{v \Delta v}{2\sqrt{a_{\text{max}} b_{\text{comfortable}}}}
\end{equation}

Here, $s_0$ is the minimum desired distance, \(T\) is the desired time headway, \(\Delta v\) is the speed difference to the vehicle ahead, and \(b_{\text{comfortable}}\) is the comfortable braking deceleration.

Vehicles approaching the crosswalk adjust their speeds to avoid collisions with pedestrians. They also assess the pedestrian’s position and crossing status to calculate safe stopping distances. If a vehicle reaches the crosswalk during a yellow or red light, it prioritizes clearing the crosswalk area to prevent blocking. To minimize collisions, vehicles dynamically adjust their speed based on proximity to pedestrians and other vehicles. A buffer zone is maintained around the crosswalk, reducing speeds when pedestrians are detected.

This comprehensive vehicle dynamics model ensures realistic traffic behavior, balancing efficient flow and safety considerations for pedestrians and other road users. The incorporation of IDM and interaction logic forms the backbone of vehicle movement in the simulation.

\subsection{Dynamic Modeling of Pedestrian Crossing Behavior}
\label{sec:pedestrian-behavior}

The pedestrian’s crossing behavior is modeled using a dynamic probability calculation that integrates \textit{patience factor, distance factor, traffic light factor,} and an added \textit{wait-for-red bias}.

The \textit{patience factor} represents a pedestrian's increasing eagerness to cross as their waiting time grows, with the rate of this growth varying based on the state of the traffic light. During a green light, patience increases slowly and is capped at an upper bound of 0.4. When the light turns yellow, patience grows at a moderately faster rate, with a slightly higher upper bound of 0.5. Finally, during a red light, patience increases rapidly, also capped at an upper bound of 0.5.

The formula for the \textit{patience factor} is shown in equation \ref{eqn:patience}, with constant factors for green, yellow, and red light equal to 20, 11.5, and 10, respectively. The waiting time is in seconds.

\begin{equation}
\label{eqn:patience}
\text { patience factor }=\min \left(\frac{\text { waiting time }}{\text { constant factor }}, \text { patience upper bound }\right)
\end{equation}

The \textit{distance factor} evaluates the proximity of the nearest vehicle to the pedestrian. When the vehicle is far, with a distance greater than four (4) meters, it increases the probability of crossing by +0.4. When the vehicle is moderately close, between two (2) and four (4) meters, it adds $ +0.1 $. However, when the vehicle is very close, with a distance of two (2) meters or less, it reduces the probability of crossing by $ - 0.4 $.

The \textit{traffic light factor} considers the state of the traffic light. A red light encourages crossing and increases the probability by $ + 0.45 $. A yellow light adds $ + 0.4 $ to the probability if vehicles are far, specifically beyond four (4) meters, but remains neutral, adding $ + 0.0 $, in other scenarios. A green light strongly discourages crossing, reducing the probability by $ -0.4 $.

An explicit \textit{wait-for-red bias} is applied during a green light and early waiting periods, defined as when the elapsed time is less than 10 seconds. This introduces an additional penalty of $ - 0.2$ to further encourage waiting for the light to turn red. This adjustment addresses the tendency of pedestrians to perceive a red light as the safest condition for crossing. The crossing threshold is calculated using equation \ref{eqn:threshold}.

\begin{equation}
\label{eqn:threshold}
\begin{aligned}
\text { crossing threshold } & =\text { base crossing threshold }+ \text { patience factor }+ \text { distance factor } \\
& + \text { traffic light factor }+ \text { wait for red bias }
\end{aligned}
\end{equation}

The \textit{crossing threshold} is bound to the range $ [0, 1] $ to ensure it represents a valid probability. If the value is negative, it is set to 0, meaning the pedestrian will not cross, while values above $1$ are capped at $1$, indicating the certainty to cross. For values within the range $[0,1]$, a random number between $0$ and $1$ is generated, and the pedestrian decides to cross if the random number is less than the crossing threshold. This probabilistic evaluation models realistic crossing behavior based on the calculated threshold.

\subsection{Monte Carlo Simulations}
\label{sec:monte-carlo}

\begin{figure}[ht]
  \centering
  \includegraphics[scale=.4]{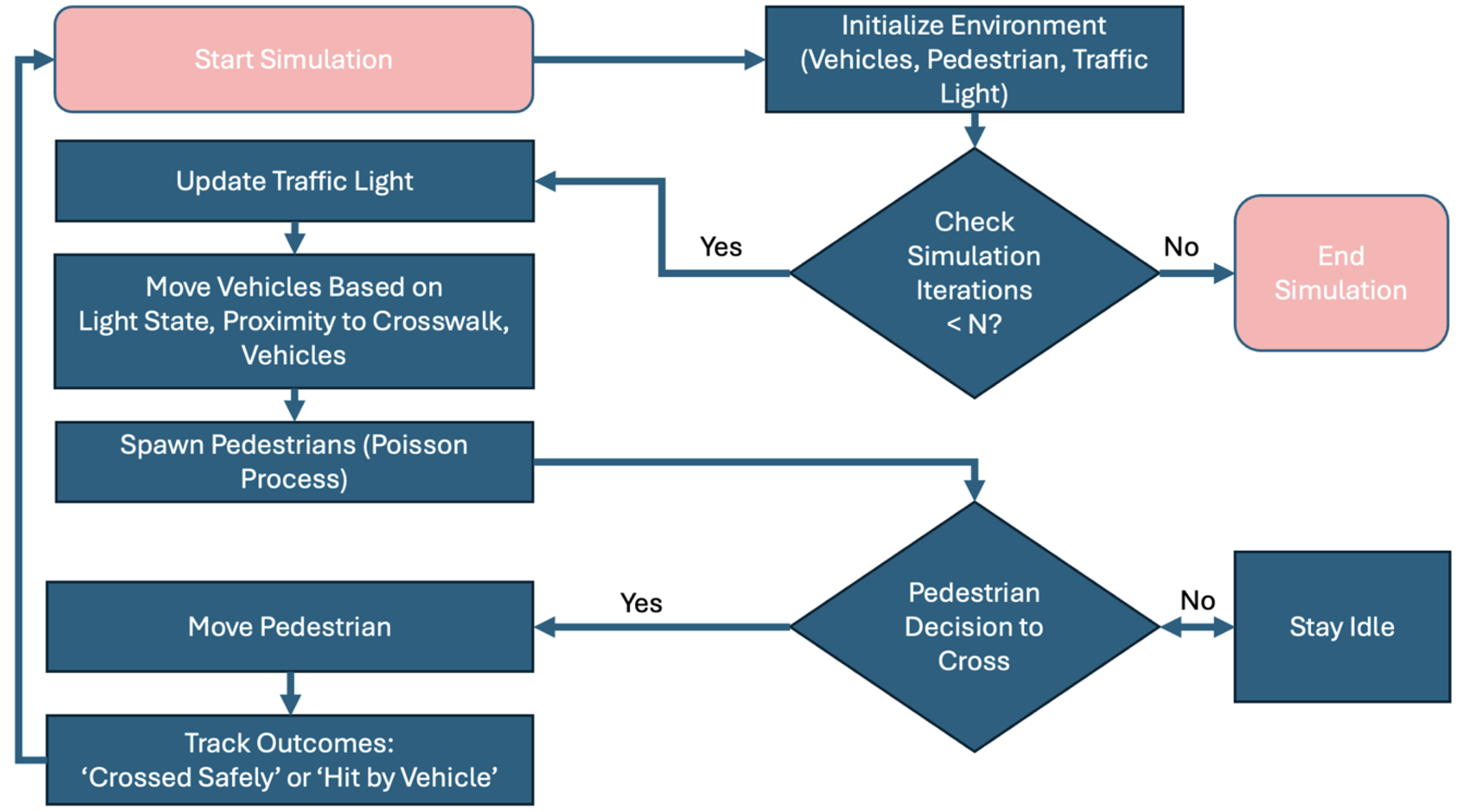}
  \caption{Monte Carlo Simulation Framework.}
  \label{fig:fig2}
\end{figure}

Monte Carlo simulations are an integral part of the proposed framework, enabling a detailed analysis of pedestrian and vehicle interactions under varying traffic conditions. By introducing randomized scenarios, this approach models the inherent variability and stochastic nature of real-world behaviors, including pedestrian decision-making and vehicle dynamics. As illustrated in Figure \ref{fig:fig2}, the simulation begins by initializing the environment, which includes vehicles, pedestrians, and the traffic light. This setup is iteratively updated across a predefined number of simulation iterations or trials, denoted as \textit{N} (in this paper, \(N = 500\) was chosen arbitrarily). At each iteration, key processes are executed: the traffic light state is updated, vehicles adjust their movements based on IDM and their proximity to the crosswalk.

The decision-making process for pedestrians, guided by dynamic probability calculations, plays a central role in the simulation. The framework evaluates whether the pedestrian decides to cross or remains idle based on real-time conditions, such as vehicle distance, waiting time, and traffic light state. If the pedestrian attempts to cross, their movement is simulated, and the outcome is tracked as either “Success” when the pedestrian safely crosses the street or “Failure” when hit by a vehicle. Simultaneously, vehicles update their positions according to IDM and dynamically responding to traffic light states and the presence of pedestrians near the crosswalk.

Figure \ref{fig:fig2} illustrates the steps of the Monte Carlo simulation framework, which supports systematic scenario testing across diverse conditions. This randomization enables the study of both high-risk and low-risk scenarios, providing robust insights into pedestrian safety under dynamic conditions. Moreover, the iterative structure allows the simulation to accommodate varying parameter configurations, ensuring scalability for large-scale urban traffic analyses. By capturing the interplay between pedestrians and vehicles in uncertain environments, this method lays the groundwork for developing adaptive traffic management solutions and digital twin models aimed at enhancing intersection safety.

The Monte Carlo simulation models high-density urban intersection scenarios by randomly initializing key variables for each trial. Pedestrian characteristics are included in the simulation, with their initial positions randomly assigned to either the top or bottom of the crosswalk. Walking speed is sampled from a normal distribution with a mean of 1.4 $m/s$ and a standard deviation of 0.2 $m/s$ to represent realistic variability among pedestrians. Their decision behavior is governed by dynamic probability calculations that consider factors such as the traffic light state, waiting time, and vehicle distance.

Vehicle dynamics are also a critical component of the simulation. Vehicles are initialized with random speeds sampled from a uniform distribution within realistic ranges, such as $10$ to $ 15\ m/s$. IDM is used to simulate car following behavior, including acceleration, deceleration, and interactions with other vehicles.

Traffic light states further influence the simulation. Transition times are randomized within defined intervals to mimic variations in real-world signal timing. Both pedestrians and vehicles dynamically adjust their behaviors based on the current traffic light state.

Each simulation trial concludes when the pedestrian either crosses the intersection safely or a collision occurs. Results are aggregated across multiple trials to evaluate crossing behaviors and safety outcomes.

\section{Results and Discussion}
\label{sec:results}

Initial test simulations were conducted with fixed initial states and deterministic pedestrian behavior to validate the model's functionality before proceeding to a probabilistic analysis using Monte Carlo simulations. These preliminary tests confirmed that pedestrians could safely cross without collisions when the traffic light for vehicles was red, assuming no other uncertainties. During yellow lights, the success rate of crossings decreased due to hesitation and variability in pedestrian behavior. The highest collision rates were observed during green lights, mirroring real-world risk patterns where vehicles have priority and pedestrian crossings are more hazardous.

\begin{figure}[ht]
  \centering
  \includegraphics[scale=.4]{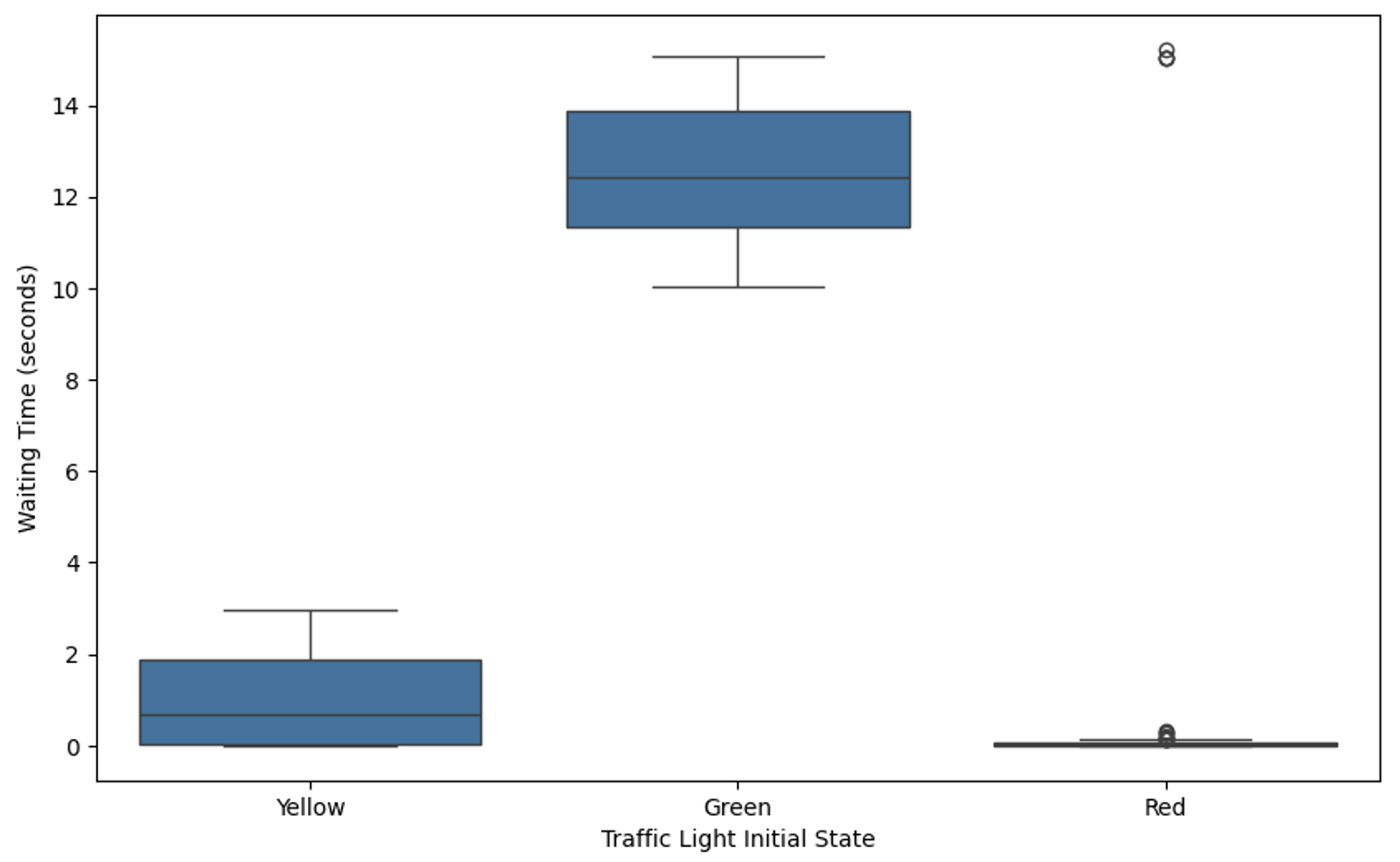}
  \caption{Waiting Time for Pedestrians by Traffic Light State.}
  \label{fig:fig3}
\end{figure}

Following the determinist tests, Monte Carlo simulations are used with 500 iterations to evaluate pedestrian behavior and crossing outcomes under diverse scenarios. The simulations accounted for key factors influencing pedestrian decision-making, including waiting time, vehicle distance, and traffic light state. Figure \ref{fig:fig3} shows pedestrian waiting times categorized by the traffic light's initial state as they wait to cross the street. Pedestrians experience the longest waiting times during green lights, with a median of around 12–14 seconds. This reflects cautious behavior as they avoid crossing under unsafe conditions and wait for a safer opportunity. During yellow lights, waiting times are shorter than green but longer than red, showing a mix of cautious and opportunistic crossing behavior depending on vehicle proximity and speed. Waiting times are minimal during red lights, as pedestrians typically cross immediately upon arrival. However, a few outliers appear when the light is red, likely representing pedestrians who arrive near the end of the red phase and must wait for the cycle to complete before crossing. This pattern aligns with the logic of safety-driven decision-making, as pedestrian behavior adapts dynamically to the traffic light state and surrounding conditions.

\begin{figure}[ht]
  \centering
  \includegraphics[scale=.4]{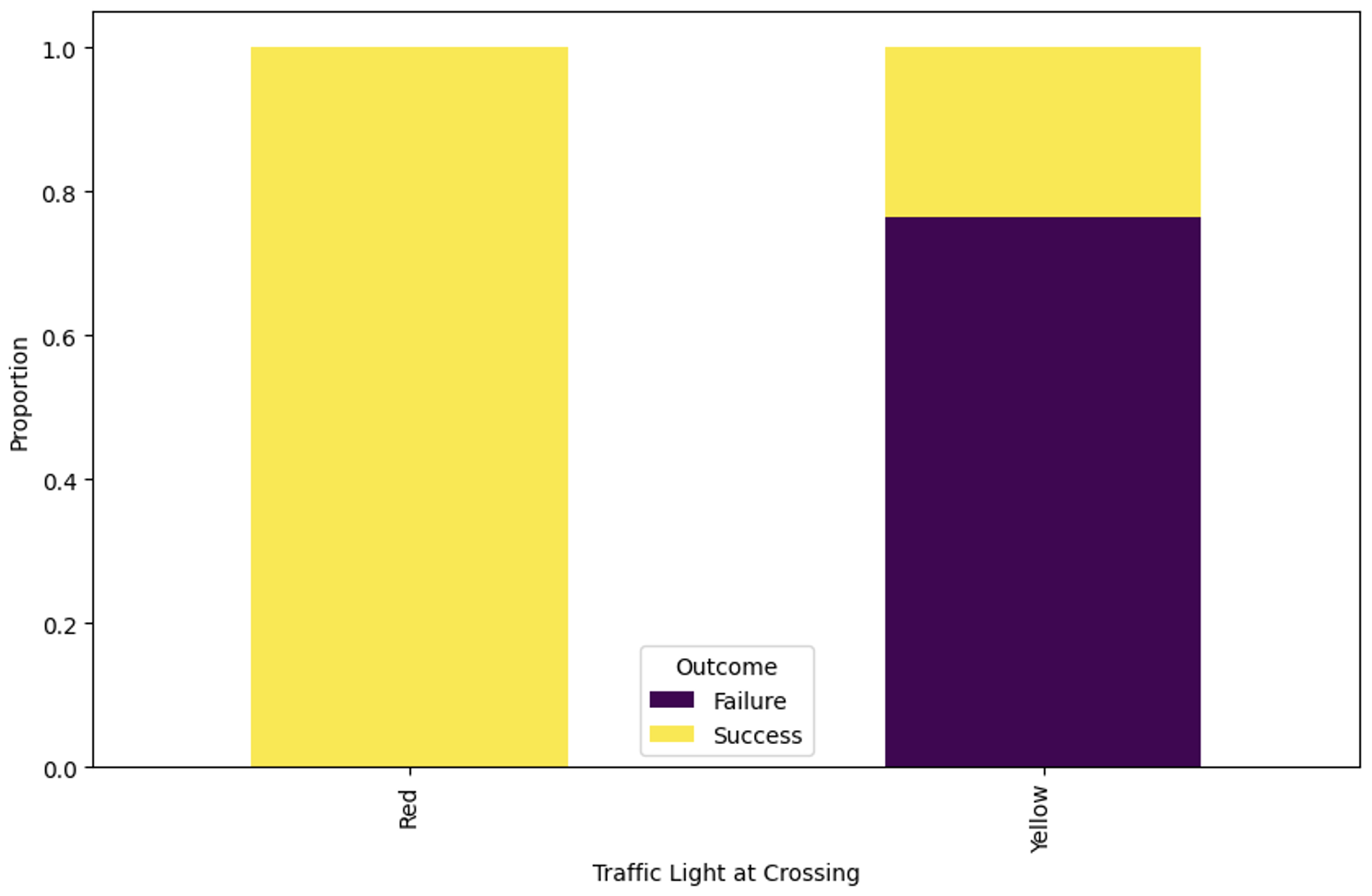}
  \caption{Success Rate by Traffic Light at Crossing.}
  \label{fig:fig4}
\end{figure}

Meanwhile, Figure \ref{fig:fig4} illustrates the proportion of successful and failed crossings based on the traffic light state at the time of crossing. Crossings during red lights are entirely successful, reflecting cautious and safe pedestrian behavior under clear traffic conditions. In contrast, crossings during yellow lights mainly result in failures, with only a small fraction being successful. These failures are likely caused by residual vehicle movement during the yellow phase or pedestrians making risky crossing decisions. The model's parameters ensure that no crossing attempts occur when the traffic light for vehicles is green, emphasizing a strong focus on safety and adherence to traffic regulations in the simulation logic.

\begin{figure}[ht]
  \centering
  \includegraphics[scale=.45]{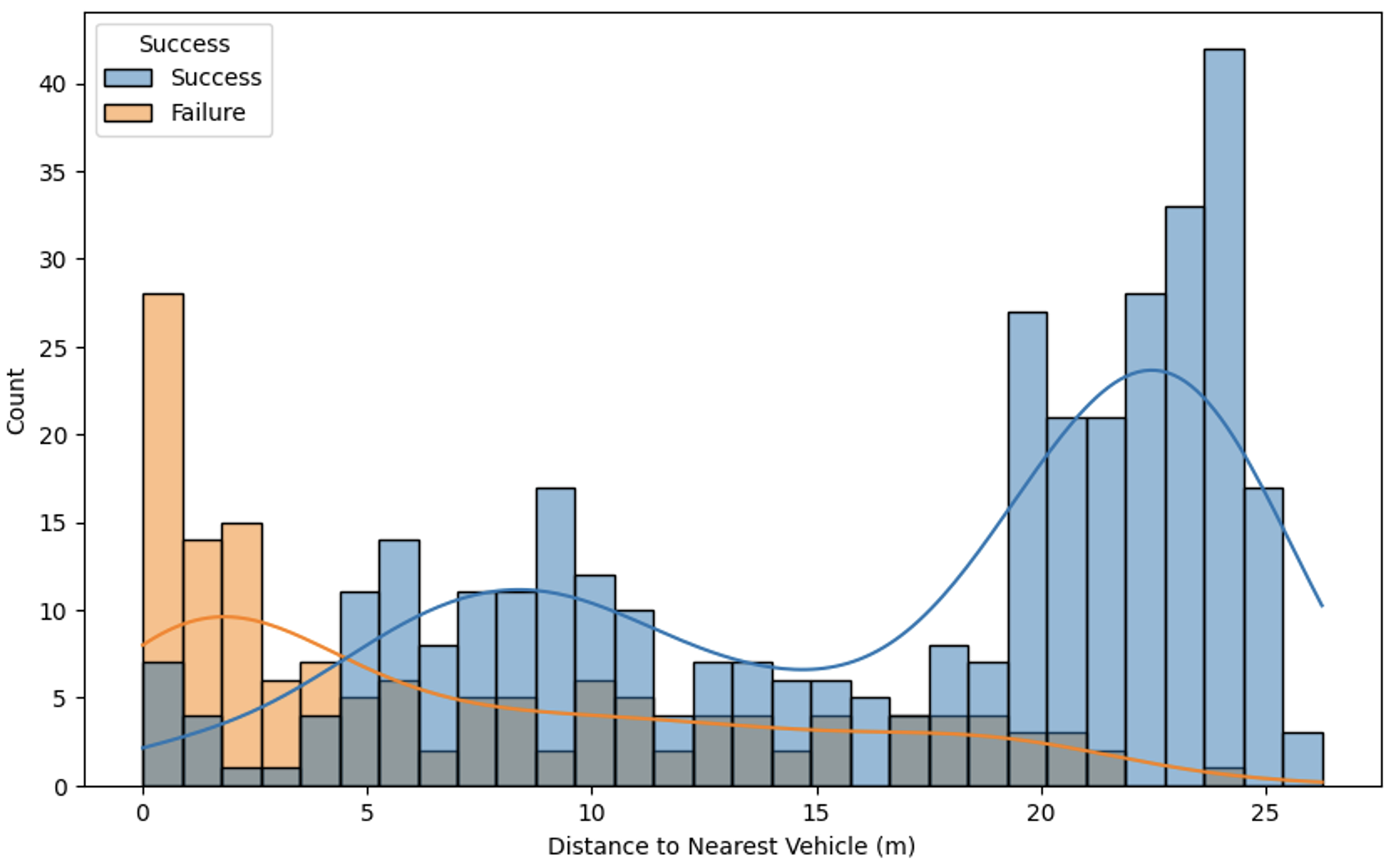}
  \caption{Distribution of Vehicle Distance and Crossing Success.}
  \label{fig:fig5}
\end{figure}

Figure \ref{fig:fig5} shows the relationship between pedestrian crossing outcomes and the distance to the nearest vehicle. Successful crossings are distributed across all vehicle distances but are most frequent when the nearest vehicle is farther away, peaking in the 20-25 meters range. Failures are highly concentrated at distances of 0-5 meters, highlighting the significant risk associated with attempting to cross when vehicles are very close. As the distance increases, failures decrease significantly, becoming much less beyond approximately 8 meters. This suggests that a distance of 8-10 meters may act as a critical threshold for safe crossing decisions, aligning with the model’s emphasis on dynamic risk assessment based on vehicle proximity.

Besides the Figures \ref{fig:fig3}--\ref{fig:fig5}, the correlation among waiting time, vehicle distance, and crossing success were also calculated. From this calculation, it has been found that success is strongly negatively correlated with pedestrian waiting time ($ - 0.81 $), indicating that shorter waiting times are strongly associated with successful crossings. Similarly, success is positively correlated with the distance to the nearest vehicle ($0.52$), showing that greater vehicle distances tend to favor successful crossings. Pedestrian waiting time and vehicle distance have a moderate negative correlation ($ - 0.55 $), suggesting that longer waits often occur when vehicles are closer. These results highlight the importance of vehicle proximity in influencing both waiting time and crossing outcomes, though the strong correlation between waiting time and success also points to waiting behavior as a critical factor in safe pedestrian decisions. While this reveals some key trends, further analysis incorporating nonlinear models or additional variables may be necessary to fully capture the dynamics of crossing behavior.

\section{Conclusion}
\label{sec:conclusion}

This study developed a pedestrian crossing simulation framework to model decision-making and collision probabilities under varying traffic conditions. The results offer valuable insights into pedestrian behavior and intersection safety, with significant implications for urban traffic management and infrastructure design.

The findings emphasize the critical role of vehicle proximity and traffic light states in shaping pedestrian decisions. Yellow lights present a higher-than-expected risk, with most crossing failures occurring during this phase, underscoring the need for dedicated pedestrian traffic lights or improved warning mechanisms to mitigate unsafe crossings. The strong positive correlation between vehicle distance and crossing success validates the model's focus on vehicle proximity as a key determinant of safe pedestrian behavior. The negative correlation between waiting time and success highlights the importance of balancing wait durations to maintain safety without encouraging risky crossings.

These results are highly dependent on the pedestrian crossing behavior model's parameters, and further refinement through Monte Carlo simulations or the inclusion of additional behavioral factors could enhance the robustness and applicability of the findings.

Practical applications include refining traffic signal designs to address risks associated with yellow light phases, such as increasing the duration of all-red intervals or implementing pedestrian-triggered signals to ensure safe crossing opportunities. Incorporating vehicle-to-everything (V2X) communication could further improve vehicle responsiveness to unpredictable pedestrian behavior, particularly in high-risk scenarios where vehicles are close.

Future research should validate these findings with real-world data, expand the model to include environmental factors (\textit{e.g., }weather, visibility, and time of day), and explore its applicability to complex urban settings, such as multi-lane intersections. Additionally, integrating real-time data into dynamic digital twins of intersections could enable adaptive traffic management systems, offering transformative potential for enhancing urban safety and efficiency.

\section*{Acknowledgment}
The authors would like to thank the Princeton University School of Engineering and Applied Science (SEAS) Innovation Grant for funding this work.

\bibliographystyle{unsrt}  
\bibliography{references}  

\end{document}